\documentclass[aps,twocolumn,showpacs,preprintnumbers,amsmath,amssymb
,nofootinbib,superscriptaddress,showkeys,prd]{revtex4-1}

\usepackage{epsfig}
\usepackage{amsmath}
\usepackage{graphicx}
\usepackage{mathptmx}      % use Times fonts if available on your TeX system
\usepackage{latexsym}
\usepackage{bm}
\usepackage{comment}

\def\slashchar#1{\setbox0=\hbox{$#1$}
   \dimen0=\wd0 \setbox1=\hbox{/} \dimen1=\wd1
   \ifdim\dimen0>\dimen1 \rlap{\hbox to \dimen0{\hfil/\hfil}} #1
   \else  \rlap{\hbox to \dimen1{\hfil$#1$\hfil}} / \fi}

\newcommand{\MeV}{\,\mathrm{MeV}}

\newcommand{\ignore}[1]{}

\begin{document}

\title{Thermal shifts, fluctuations, and missing states}\thanks{Talk
    presented by E.R.A. at the YSTAR2016 meeting ``Excited Hyperons in QCD
    Thermodynamics at Freeze-Out'', 16-17 November 2016, Thomas
    Jefferson National Accelerator Facility, Newport News, VA. }

\author{Enrique \surname{Ruiz Arriola}}
\email{earriola@ugr.es}

\affiliation{Departamento de F\'{\i}sica At\'omica, Molecular y Nuclear and
  Instituto Carlos I de F\'{\i}sica Te\'orica y Computacional \\ Universidad
  de Granada, E-18071 Granada, Spain.}

\author{Wojciech Broniowski}
\email{Wojciech.Broniowski@ifj.edu.pl}
\affiliation{Institute of Physics, Jan Kochanowski University, 25-406 Kielce, Poland}
\affiliation{The H. Niewodnicza\'nski Institute of Nuclear Physics, Polish Academy of Sciences, PL-31342 Cracow, Poland}

\author{Eugenio Meg\'{\i}as}
\email{emegias@mppmu.mpg.de}

\affiliation{Max-Planck-Institut f\"ur Physik (Werner-Heisenberg-Institut), F\"ohringer Ring 6, D-80805 Munich, Germany}

\affiliation{Departamento de F\'{\i}sica Te\'orica, Universidad del
  Pa\'{\i}s Vasco UPV/EHU, Apartado 644, 48080 Bilbao, Spain}

\author{Lorenzo L. Salcedo}
\email{salcedo@ugr.es}

\affiliation{Departamento de F\'{\i}sica At\'omica, Molecular y Nuclear and
  Instituto Carlos I de F\'{\i}sica Te\'orica y Computacional \\ Universidad
  de Granada, E-18071 Granada, Spain.}

\date{17 December 2016}

\begin{abstract}
\medskip
Thermal shifts and fluctuations at finite temperature below the
deconfinement crossover from hadronic matter to the quark-gluon plasma
provide a viable way to search for missing states with given quantum
numbers in the hadronic spectrum. We analyze three realizations of the
hadron resonance gas model in the light quark ($uds$) sector: the
states from the Particle Data Group tables with or without width and
from the Relativized Quark Model. We elaborate on the meaning of
hadronic completeness and thermodynamical equivalence on the light of
lattice QCD trace anomaly, heavy quark entropy shift and
baryon,charge and strangeness susceptibilities.
\end{abstract}



\maketitle

\section{Introduction}

The concept of missing states in QCD is intimately related to the
completeness of the hadronic spectrum.  The issue was anticipated by
Hagedorn in the mid 60's~\cite{Hagedorn:1965st} when analyzing the
mass-level density $\rho(M)$ and {\it predicting} the bulk of states
at higher masses, which later on were experimentally confirmed.  This
also implies that the states may be {\it counted} one by one (and
hence ordered) by, say, the cumulative number of states function,
\begin{eqnarray}
N(M) = \sum_n g_n \theta (M-M_n) \, , 
\label{eq.count}
\end{eqnarray}
with $g_n$ the total degeneracy and $\rho(M)=dN(M)/dM$.  Updated
analyses of the Hagedorn hypothesis may be found
in~\cite{Broniowski:2000bj,Broniowski:2004yh}. The function $N(M)$
assumes integer values, and the best mass resolution is $\Delta M =
\min_n (M_{n+1}-M_n)$. For bound states, where the spectrum is
discrete, this is a well defined procedure. In the continuum, this can
only be done by putting the system in a box with finite but
sufficiently large volume which acts as an infrared cut-off $ V^{1/3}
\Delta M \gg 1$. The ultraviolet cut-off is the maximum mass $M_{\rm
  max}$ in Eq.~(\ref{eq.count}).

The commonly accepted reference for hadronic states is the Particle
Data Group (PDG) table~\cite{Olive:2016xmw}, a compilation reflecting
a consensus in the particle physics community and which grades states
*,**,***, and ****, according to the growing confidence in their
existence, respectively.  Global features of the hadronic spectrum may
depend on whether we decide to promote or demote their significance,
according to some theoretical prejudice. Of course, we expect the PDG
hadronic states to have a one-to-one correspondence with colour
neutral eigenstates of the QCD Hamiltonian; indeed, ground and some
excited states have been determined on the
lattice~\cite{Edwards:2012fx}. For hadronic states with only light
($uds$) quarks the maximum mass, $M_{\rm max}$, recorded by PDG is
around $2.5~{\rm GeV}$ for mesons and baryons, hence currently
$N(M_{\rm max}) \sim 2.5 \times 10^3$. So far, the states listed by
PDG echo the standard quark model classification for mesons ($\bar q
q$) and baryons ($qqq$). Because of this feature, it will be pertinent
to consider also the Relativized Quark Model (RQM) for
mesons~\cite{Godfrey:1985xj} and Baryons~\cite{Capstick:1986bm}, as
first done for $N(M)$ in~\cite{Broniowski:2000hd}~(Fig.~9).  The
remarkable coincidence $N_{\rm PDG} (M) \sim N_{\rm RQM} (M)$ up to
$M_{\rm max}$ for {\it both} mesons and baryons has been shown in
Ref.~\cite{Arriola:2014bfa}. The so-called ``further states'' may or
may not be confirmed or expected and have not been clearly regarded by
the PDG as identified, although they could be exotic tetraquarks,
$\bar q q \bar q q $, pentaquarks, $\bar q qqqq$, glueballs $gg$,
$ggg$ or hybrids $\bar q qg$~\cite{Briceno:2015rlt}.

In this contribution we analyze thermodynamic measures (various
susceptibilities) which are sensitive to missing states.  The setup
corresponds to heating up the vacuum without dissolving its
constituent hadrons into quarks and gluons and testing quark-hadron
duality at temperatures below $T_c \sim 150 {\rm MeV}$.  Obviously,
such a framework is inefficient for individual states, but becomes
competitive if globally a relatively large number of states are
missing.  As reported in~\cite{Majumder:2010ik,Broniowski:2016hvt} the
Hagedorn conjectured behaviour of $N(M) \sim A e^{M/T_H}$ for $M >
M_{\rm max}$ may influence the results close to $T_c$, at temperatures
above $T>140$~MeV.  According to~\cite{Broniowski:2016hvt}, there is
not much room for such states in the one-body observables, where they
would spoil the agreement with the lattice data, unless suitable
repulsion between states is simultaneously incorporated.  Here will
make no attempt to complete the spectrum beyond $M_{\rm max}$.

\section{Prehistory of Missing states}

The use of thermodynamical arguments to characterize the existence of
missing states is a rather old subject which goes back to the early
beginnings of the kinetic theory of gases and the equipartition
theorem. In its most general form it states that every degree of
freedom contributes to the mean energy with $\frac12 k_B
T$.\footnote{The story around this principle illustrates many of the
  issues under discussion, including the contribution of an anonymous
  referee~\cite{brush1967foundations}
  D. Bernoulli~\cite{bernoulli1738hydrodynamica} was the first who
  found in 1738 that the Boyle-Mariotte, Gay-Lussac, and Charles
  equations could be unified and understood by means of Newton's
  equations and in statistical terms. His work was forgotten, and only
  in 1845 J. J. Waterston submitted a paper to the Philosophical
  Transactions of the Royal Society (PTRS) which was rejected with a
  remark ``The paper is nothing but nonsense, unfit even for reading
  before the Society''. Hence this work was also ignored. Maxwell, in
  1859, managed to publish the case of rigid molecules, and Boltzmann
  generalized it in 1868 to its modern form including rotational and
  vibrational degrees of freedom. Lord Rayleigh in 1895, found by
  chance Waterston's paper in the archives and decided to publish it
  in PTRS twelve years after Waterston's death with a commentary:
  ``had he put forward his investigation as a development of Bernoulli
  a referee migh have hesitated to call it nonsense. It is probable
  that Waterston was unaquainted with his work.''.}.  Therefore $\bar
E = N \nu k_B T/2$, where $\nu$ is the total number of degrees of
freedom.  Generally, $\nu = \nu_{\rm translation}+ \nu_{\rm
  vibration}+ \nu_{\rm rotation} $ and the molar specific heat is
$c_V/R= \nu/2 $. A major obstacle at the time was pointed out by
J. C. Maxwell in 1860 in connection to the specific heat of the
diatomic gas such as, e.g., $H_2$, where a priori the total number of
degrees of freedom is $\nu= 3_{\rm trans}+ 2_{\rm vib}+ 2_{\rm rot} =
7$. This would imply $c_V/R = 7/2$, whereas experimentally at room
temperature one has $c_V/R \approx 5/2 $.  This is because the
vibrational degrees of freedom are not active due to high excitation
energy, and become visible only as the dissociation temperature of
$\sim 3200$~K is approached. Likewise, as $T$ is decreased, the
rotational degrees of freedom are also frozen and below $\sim 70$~K,
$c_V/R = 3/2$, as for the monoatomic molecules.

In modern terms the ``freezing'' of degrees of freedom is related to
the quantization of energy levels for the Hamiltonian $H \Psi_n = E_n
\Psi_n$ with energy eigenvalues above the temperature, $ E_n > T$,
contributing negligibly to the partition function
\begin{eqnarray}
Z= {\rm Tr} e^{-H/T}= \sum_n e^{-E_n/T}. \label{eq:Z}
\end{eqnarray}
In QCD, the quantized energy levels are the masses of the existing
hadronic states and, like in the Maxwell argument, the states which are not
activated when $M_n > T$ do not contribute.

\section{Completeness of the hadron spectrum}

Completeness of the listed PDG states~\cite{Olive:2016xmw} is a
subtle issue.  On the one hand they are mapped into the $\bar q q $
and $qqq$ quark model states. On the other hand, most reported states
are not stable particles but resonances produced
as intermediate steps in a scattering process.

With a finite lifetime $\tau_R$, they are characterized by a mass
distribution $\rho_R (M)$, with a central value $M_R$ and a width
$\Gamma_R \sim \hbar/\tau_R$.  From a rigorous point of view 
resonances are poles of the {\it exact} amplitude in the second
Riemann sheet in the complex $s$ plane at $s = M^2 - i M \Gamma$. For
multichannel scattering with $N$ channels one has $2^N$ Riemann
sheets, depending on which cuts have been crossed (see, e.g.,
\cite{Nieves:2001wt,GarciaRecio:2002td} for discussions in the
meson-baryon $S=0,-1$ sectors). Despite the rigor of these
definitions, complex energies are not directly measured. An analytic
continuation of a phenomenological and approximate scattering
amplitude, taking into account a process dependent background, is
needed and the arbitrariness grows with the width of the
resonance~\cite{Ciulli:1975sm} (see e.g. for the specific $0^{++}$
case~\cite{Caprini:2016uxy}). On average, most of the resonances
listed by PDG~\cite{Olive:2016xmw} can be regarded as narrow, since
one finds $ \langle \Gamma_R/M_R \rangle = 0.12(8) $ {\it both} for
mesons and baryons~\cite{Arriola:2011en,Masjuan:2012gc}, a fact
numerically consistent with the large $N_c$ theoretical expectation
$\Gamma_R/M_R = {\cal O} (N_c^{-1})$~\cite{Witten:1979kh}. In the
Hamiltonian picture, resonances are identified as the so-called Gamow
states and are not normalizable in the usual Hilbert space, as they
are not conventional irreducible representations of the Poincar\`e
group~\cite{Bohm:2004zi}. The completeness relation involves bound
states and the continuum, which can be rewritten as a discrete sum of
the Gamow states and a remainder~\cite{Berggren:1968zz}.

The meaning of completeness is fairly clear within a given Hilbert
space ${\cal H}$ with specified degrees of freedom when only
bound states are possible. For instance, if we restrict ourselves to the meson
$(\bar q q)$ or baryon $(qqq)$ sectors, such as in 
RQM~\cite{Godfrey:1985xj,Capstick:1986bm}, we can diagonalize the
$\bar q q$ and $qqq$ Hamiltonians with confining potentials in a given
already complete basis, which is truncated but large enough that states with $M_n \le
M_{\rm max}$ converge. Thus we write
\begin{eqnarray}
{\cal H}_{\rm RQM} = {\cal H}_{\bar q q } \oplus {\cal H}_{qqq} \oplus
{\cal H}_{\bar q \bar q \bar q}
\end{eqnarray}
Within this framework, hadrons are stable, extended, and composite
particles. This is explicitly illustrated by the virial relations in
the massless quark limit~\cite{Arriola:2014bfa} $ M_{\bar q q} = 2
\sigma \langle r \rangle_{\bar q q} $ and $ M_{qqq} = N_c \sigma
\langle r \rangle_{qq} $, which shows that hadrons are larger the
heavier they become. Many of these states may decay by strong
processes, such as $\rho \to 2 \pi$ or $\Delta \to N \pi$, where 
a coupling to the continuum is needed by incorporating the ${\cal
  H}_{\bar q q \bar q q } $ and ${\cal H}_{\bar q q qqq}$ Fock state.
As a result, the pole mass is shifted into the complex plane $ M \to M +
\Delta M - i \Gamma/2 $. The mass-shift $\Delta M \sim \Gamma$ depends
parametrically on the coupling to the continuum $\Delta M \sim \Gamma$
so that in the large $N_c$ limit, $\Delta M /M = {\cal O}
(N_c^{-1})$~\cite{Masjuan:2012sk}. 

On the lattice, hadrons are constructed as interpolating fields in a
finite-volume box. Completeness proceeds along similar lines, with the
important modification that resonances are characterized by volume-independent 
and real mass shifts. The connection to physical
resonances in the complex energy plane requires also analytical
extrapolation (for a review see, e.g., \cite{Briceno:2014tqa}).

\section{Thermodynamic equivalence}

Be it the PDG~\cite{Olive:2016xmw}, RQM~\cite{Godfrey:1985xj,Capstick:1986bm}, or the lattice excited
QCD~\cite{Edwards:2012fx}, the partition function can be constructed
from the (complete) energy localized colour neutral eigenstates,
Eq.~(\ref{eq:Z}). The lattice at finite temperatures, or the ultrarelativistic heavy ions collisions, generate global colour neutral configurations which
along the crossover are expected to delocalize. Most of the emerging
physical quark-hadron duality picture has to do with the
thermodynamical equivalence of different approaches.

According to the quantum virial expansion~\cite{Dashen:1969ep} one can
compute the partition function from the knowledge of the $S$-matrix in
the complete Hilbert space, i.e., involving all possible processes
with any number of elementary particles in both the initial and final
states, $n \to m$. In practice, hadrons have been taken as the
building blocks in this approach, which for obvious practical reasons
has never been taken beyond the $2 \to 2$ reactions, where the
corresponding phase shifts are involved. In the case of narrow
resonances one can replace the total contribution entering in terms of
phase shifts by the resonance itself~\cite{Dashen:1974jw}, whereby the
resonance can be assumed to be elementary and
point-like~\cite{Dashen:1974yy}. The result conforms to the Hadron
Resonance Gas (HRG) as initially proposed by
Hagedorn~\cite{Hagedorn:1965st}.  This provides the formal basis for
modern HRG calculations using the PDG compilation. As mentioned above,
most states entering the HRG are resonances with a given width,
$\Gamma$. Therefore we will also consider the effect of smearing the
mass distribution
according to the replacement
\begin{eqnarray}
\sum_R F(m_R^2) \to \int d\mu^2 F(\mu^2) \Delta_\Gamma ( \mu^2
-m_R^2)
\end{eqnarray}
for an observable $F(\mu^2)$.\footnote{Ideally the profile function
  should be determined from the scattering
  phase-shift~~\cite{Dashen:1974jw}, which displays
  cancellations~\cite{Venugopalan:1992hy,Broniowski:2015oha} and
  irrelevance of some weakly bound states~\cite{Arriola:2015gra} but
  it is not always available. Here we take a simple normalized
  Gaussian profile distribution. A Breit-Wigner representation works
  well around the resonance, but it has very long tails which do not
  faithfully represent the background. An upper bound for the error is
  to use the half-width rule~\cite{Arriola:2011en,Arriola:2012vk}
  according to which PDG masses are varied within half the width,
  i.e. taking $M_R \pm \Gamma_R/2$. We do not use this large $N_c$
  motivated prescription here as we feel that it largely overestimates
  the uncertainties for {\it all resonances} in the $N_c=3$ world.}

However, the elementary constituents are both quarks and gluons. A
different derivation proceeds along chiral quark-gluon models with a
quantum and local Polyakov
loop~\cite{Megias:2004hj,Megias:2006bn}.\footnote{This is unlike the
  more popular PNJL model~\cite{Fukushima:2003fw,Ratti:2005jh}, where
  the quantum and local nature of $\Omega (\vec x) $ is ignored, thus 
  introducing an undesirable group coordinates dependence.  In
  addition, in PNJL the Polyakov loop in the adjoint representation is
  not quenched, contradicting lattice calculations.} The action
corresponds to creating, e.g. a quark at location $\vec x$ and momentum
$\vec p$ in the medium
\begin{eqnarray}
e^{-E(\vec p)/T} \Omega(\vec x)^\dagger,
\end{eqnarray}
where in the static gauge $\Omega(\vec x)= e^{i g A_4 (\vec
  x)/T}$. Consequently, the total action can be separated into
different quark and gluon sectors according to the low temperature
partonic expansion around the
vacuum~\cite{RuizArriola:2012wd,Arriola:2014bfa}
\begin{eqnarray}
Z= Z_0+ Z_{\bar q q }+ Z_{qqq} + Z_{\bar q \bar q \bar q} + \dots \sim
Z_{\rm RQM}.
\end{eqnarray}
Subsequent hadronization of $\bar q q$ and $qqq$ states uses the
cluster properties of the Polyakov loop correlator and group
properties of the Haar measure, as well as the quantum, composite and
extended nature of hadronic states. One appealing feature of this
``microscopic'' derivation of the HRG is the counting of states
according to the quark model for the lowest Fock state components, but
ambiguities arise when a given colour neutral multiquark state admits
a separation into colour neutral irreducible
subsystems~\cite{Arriola:2014bfa,Megias:2013xaa}. We take this result
as our justification to use RQM.\footnote{Bound state masses are
  shifted when coupled to the continuum, so if we take a simple 
  average estimate $\langle \Delta M/M \rangle_{\rm RQM} \sim \langle
  \Gamma/M \rangle_{\rm PDG} \sim 0.12(8)$.  This roughly corresponds
  to take $5\%-20\%$ uncertainty in $T$.}

The fact that we use thermodynamic quantities to make a quantitative
comparison does not sidestep the problem of discriminating different
spectra. The best example is provided by a direct comparison of
HRG using either PDG or 
RQM~\cite{Godfrey:1985xj,Capstick:1986bm} in terms of the trace
anomaly, ${\cal A}(T) \equiv (\epsilon-3 P)/T^4$ which are hardly
indistinguishable within the lattice QCD uncertainties from the
WB~\cite{Borsanyi:2013bia} and HotQCD~\cite{Bazavov:2014pvz}
collaborations (see Fig.1 of Ref.~\cite{Arriola:2014bfa}).
As already mentioned the states with $M > M_{\rm max}$ with an
exponential Hagedorn distribution are relevant below
$T_c$~\cite{Broniowski:2016hvt} only at $T >140$~MeV, and their
contribution may be overcome with repulsive effects. Actually, the
volume effects are expected to play a significant role; the excluded
volume exceeds the total volume around $T \lesssim T_c$ (see Fig.~9 of
Ref.~\cite{Arriola:2014bfa}).

\section{Thermal shifts}

The idea of thermal shifts is to study the change of thermodynamic
quantities under the presence of local external sources. This looks
very much like adding an impurity to a macroscopic system or adding a
grain of salt to a bunch of snow. By looking a these thermal shift we
may also assess a possible existence of missing states.  An
interesting hadronic example is provided by the free energy shift
caused by a heavy quark placed in a hot medium with vacuum quantum
numbers, which corresponds to a ratio of partition functions which can
be identified with the Polyakov loop expectation value. This free energy shift is
ambiguous and hence it is better to deal with the corresponding
entropy shift and the specific heat, which are directly measurable
quantities. A hadronic representation of Polyakov loop and its entropy
has been analyzed~\cite{Megias:2012kb,Megias:2012hk}. The implications of thermal
shifts due to a heavy source or a heavy $Q \bar Q$ pair located at a
fixed distance $r$ at the hadronic level has recently been
considered in~\cite{Megias:2016bhk,RuizArriola:2016ift,Megias:2016onb}.

\begin{figure}[tbc]
\begin{center}
\epsfig{figure=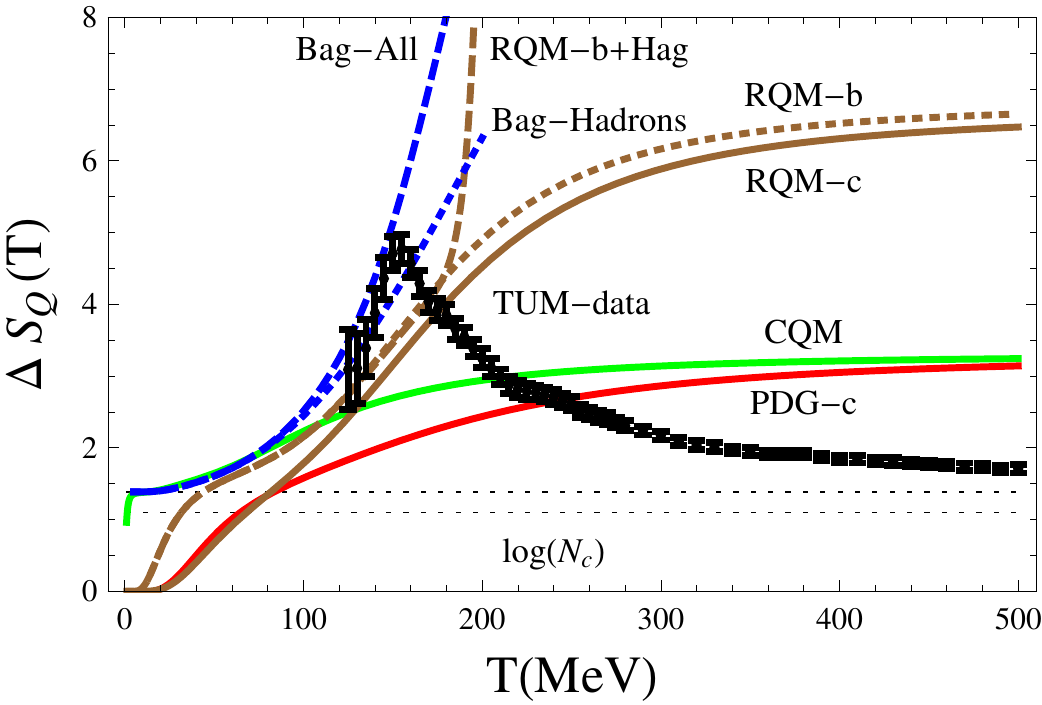,height=6cm,width=8cm}
\end{center}
\caption{The entropy as a function of the temperature.  We show
  results from various hadronic models: the bag model including all
  ($Q\bar q$, $Q qq$ and $Q\bar qg$) states and just hadrons, the RQM
   with one $c$- or $b$-quark and the PDG states with one
  $c$-quark. The Hagedorn extrapolation of the $b$-spectrum is also
  displayed. We also plot the CQM with $uds$ quarks and constituent
  mass $M=300 \MeV$ and the bare $m_u=2.5 \MeV $, $m_d= 5 \MeV$, $m_s=
  95 \MeV$ masses. Horizontal lines mark $\Delta S_Q(0) = \log 2 N_f$,
with $N_f=2$ the number of light degenerate flavours, 
  and $\Delta S_Q(\infty)= \log(N_c)$. Lattice data for 2+1
flavours are taken from
  Ref. \cite{Bazavov:2016uvm}.}
\label{fig:entropy-models}
\end{figure}

Fig.~\ref{fig:entropy-models} from Ref.~\cite{Megias:2016bhk} makes a
good case for different categories of missing charm or bottom
states. On the one hand the PDG is clearly insufficient to describe
the entropy shift. So, we clearly miss higher mass states. Guided by
the thermodynamic equivalence of PDG and RQM in the $uds$
sector~\cite{Arriola:2014bfa}, we may complete the PDG spectrum
using the RQM in the c- or b- sectors. As we see there is a big
improvement and, moreover, the change when going from c to b is
sufficiently small. Nonetheless, we have still missing states, a
feature that is not mended when extending the spectrum a la
Hagedorn. When a Bag model with the heavy source located at the center
is considered for singly heavy hadrons $Q \bar q$, $Q q q$ and a
hybrid $Q \bar q g$ the TUM lattice data are well reproduced.

\section{Fluctuations}

The connection between fluctuations and the abundance of hadronic
resonances was pointed out by Jeon and Koch~\cite{Jeon:1999gr}, who
later~\cite{Jeon:2000wg} proposed it as a signal for the Quark-Gluon
Plasma formation  from the partition
function (for pedagogical reviews see,
e.g.,~\cite{Koch:2008ia,Asakawa:2015ybt}). Implications for heavy ion collisions are reviewed
in~\cite{Braun-Munzinger:2015hba}. In~\cite{Bazavov:2014xya}, the
event-by-event statistical analysis of ultrarelativistic heavy
ions-collisions was compared to the HRG with a given chemical
potential. Of course, any mismatch in this kind of analyses suggests
missing resonances. Here we are concerned with the simplest vacuum
zero density case. Actually, some authors have understood the
significance of fluctuations as a possible hint of missing
states~\cite{ManLo:2016pgd}.

Fluctuations of conserved charges, i.e. fulfilling $[Q_A,H]=0$, are a
way of selecting given quantum numbers~\cite{Asakawa:2015ybt} and
become particularly simple in terms of the grand-canonical partition
function which is given by
\begin{eqnarray}
Z= {\rm Tr} e^{- (H - \sum_A \mu_A  Q_A )/T} \qquad \Omega = - T \log Z. 
\end{eqnarray}  
with $\Omega$ the corresponding potential. One then gets
\begin{equation}
-\frac{\partial \Omega}{\partial \mu_A} = \langle Q_A \rangle_T ,
\quad 
- T \frac{\partial^2 \Omega}{\partial \mu_A \partial \mu_B} = 
\langle \Delta Q_A \Delta Q_B \rangle_T  
,
\end{equation}
where $\Delta Q_A = Q_A - \langle Q_A \rangle_T$. In the $uds$ sector the
only conserved charges are the electric charge $Q$, the baryon charge $B$ and
the strangeness $S$, which is equivalent to the number of $u$, $d$, and $s$
quarks. We consider the hot vacuum (no chemical potential) $ \langle B
\rangle_T = \langle Q \rangle_T = \langle S \rangle_T = 0$.

For $N_f=2+1$, fluctuations have been computed on the lattice by the
WB~\cite{Borsanyi:2011sw} and HotQCD~\cite{Bazavov:2012jq}
collaborations with the high temperature asymptotic limits 
\begin{eqnarray}
\chi_{BB}(T) &=& V^{-1}\langle B^2 \rangle_T \to \frac1{N_c} \\  
   \chi_{QQ} (T) &=&  V^{-1} \langle Q^2 \rangle_T \to \sum_{i=1}^{N_f} q_i^2 \\
  \chi_{SS}(T) &=& V^{-1} \langle S^2 \rangle_T \to 1, 
\end{eqnarray}
where $(q_u,q_d,q_s, \dots ) = (2/3,-1/3,-1/3, \dots)$. Higher order
cumulants, such as skewness and kurtosis originally analyzed in
Ref.~\cite{Ejiri:2005wq}, have also recently been computed more
accurately~\cite{Bellwied:2015lba}, but we do not discuss them here.

In the hadron resonance model, the charges are carried by various species of
hadrons, $Q_A = \sum_i q^{(i)}_A N_i$, where $N_i$ is the number of hadrons of
type $i$, hence
\begin{equation}
\langle \Delta Q_A  \Delta Q_B \rangle_T 
= \sum_{i,j}  q^{(i)}_A q^{(j)}_B \langle \Delta N_i \Delta N_j \rangle_T
.
\end{equation}
The average number of hadrons is
\begin{equation}\begin{split}
\langle N_i \rangle_T &=
V \int \frac{d^3 k}{(2\pi)^3} \frac{g_i}{e^{E_{k,i} /T} + \eta_i} \nonumber \\ 
&= \frac{VT^3}{2\pi^2} \sum_{n=1}^\infty g_i \frac{(-\eta_i)^{n+1}}{n}
  \left( \frac{M_i}{T} \right)^2 K_2(n M_i/T)
\end{split}\end{equation}
where $E_{k,i}= \sqrt{M_i^2 + k^2}$, $g_i$ is the degeneracy and $\eta_i=\mp
1$ for bosons/fermions respectively. In practice the Boltzmann approximation
(i.e., just keeping $n=1$) is sufficient. Regarding the fluctuations, 
since the different species are uncorrelated
$\langle \Delta n_\alpha \Delta n_\beta \rangle_T = \delta_{\alpha\beta}
\langle n_\alpha \rangle_T (1 - \eta_\alpha \langle n_\alpha \rangle_T )$, for
the occupation numbers. Since $\langle n_\alpha \rangle_T \ll 1$,
\begin{eqnarray}
\langle \Delta Q_A \Delta Q_B \rangle_T \approx
\sum_{i} q_A^{(i)} q_B^{(i)} \langle N_i \rangle_T.
\end{eqnarray}

\begin{figure*}
\epsfig{figure=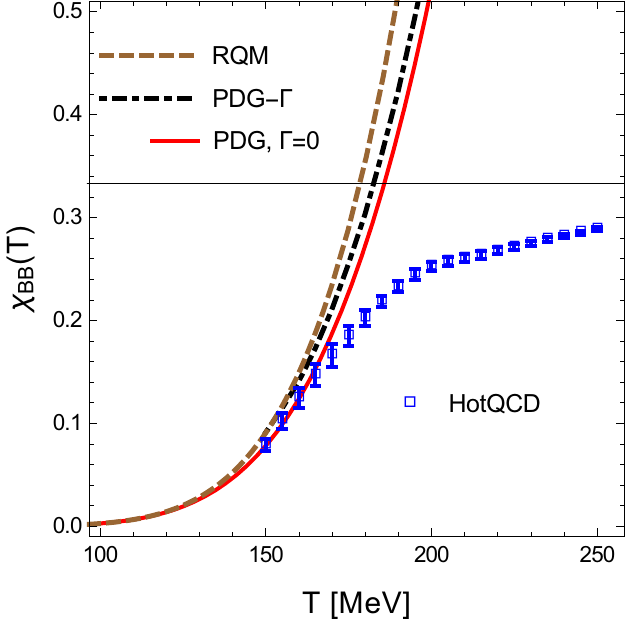,width=5.7cm,angle=0}
\epsfig{figure=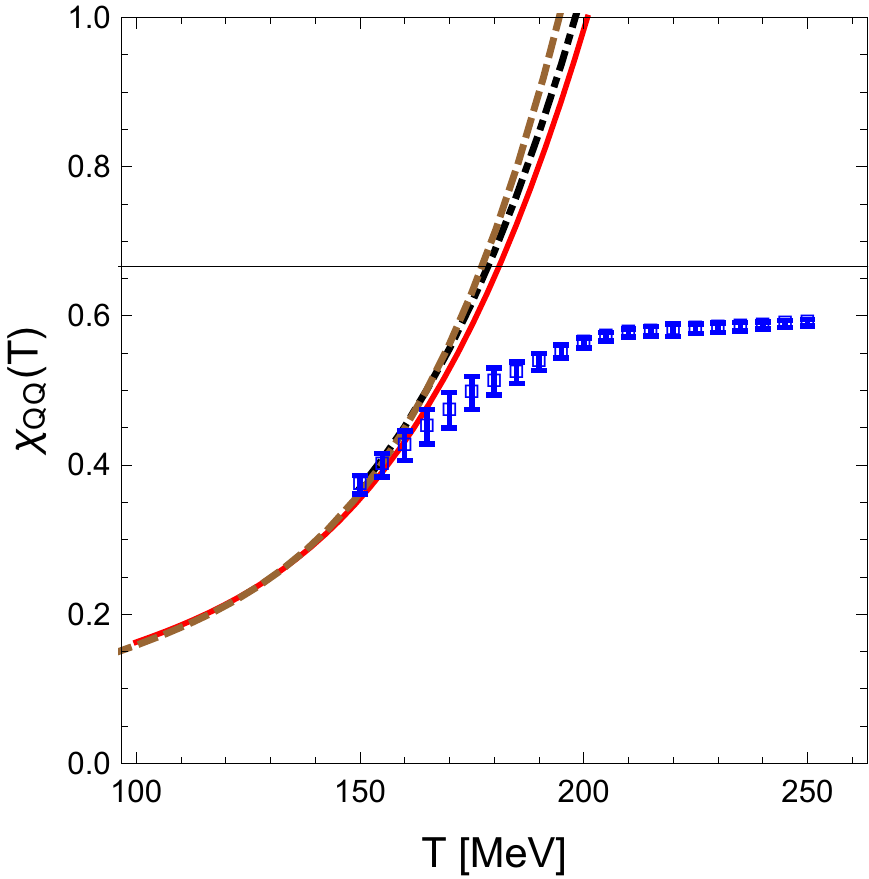,width=5.7cm,angle=0}
\epsfig{figure=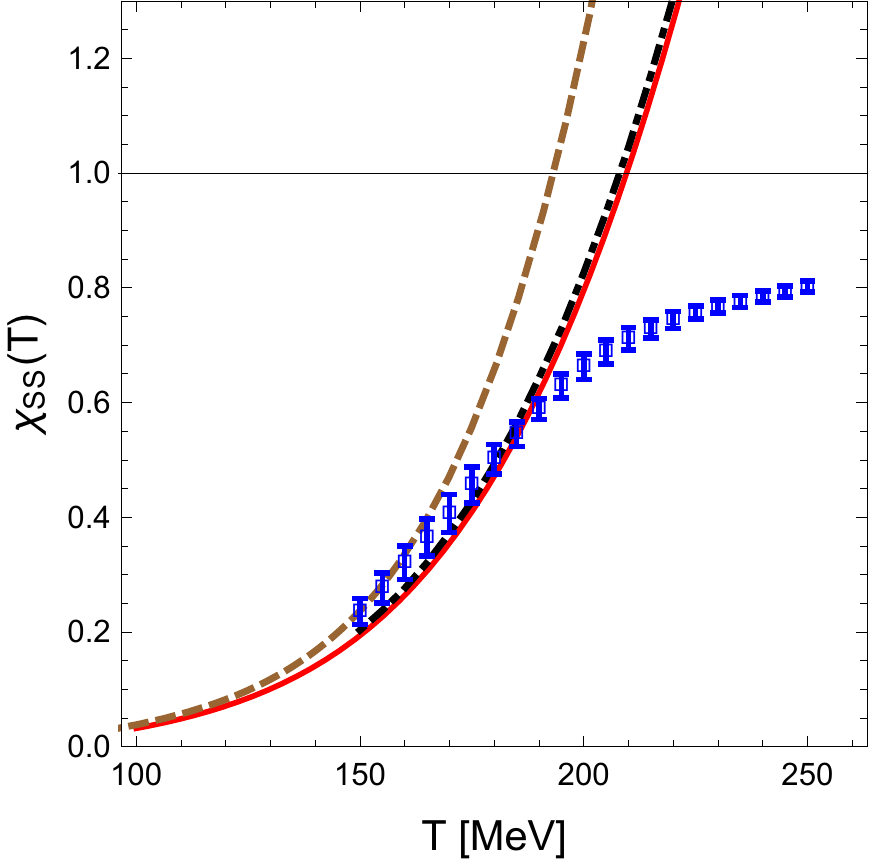,width=5.7cm,angle=0}\\

\epsfig{figure=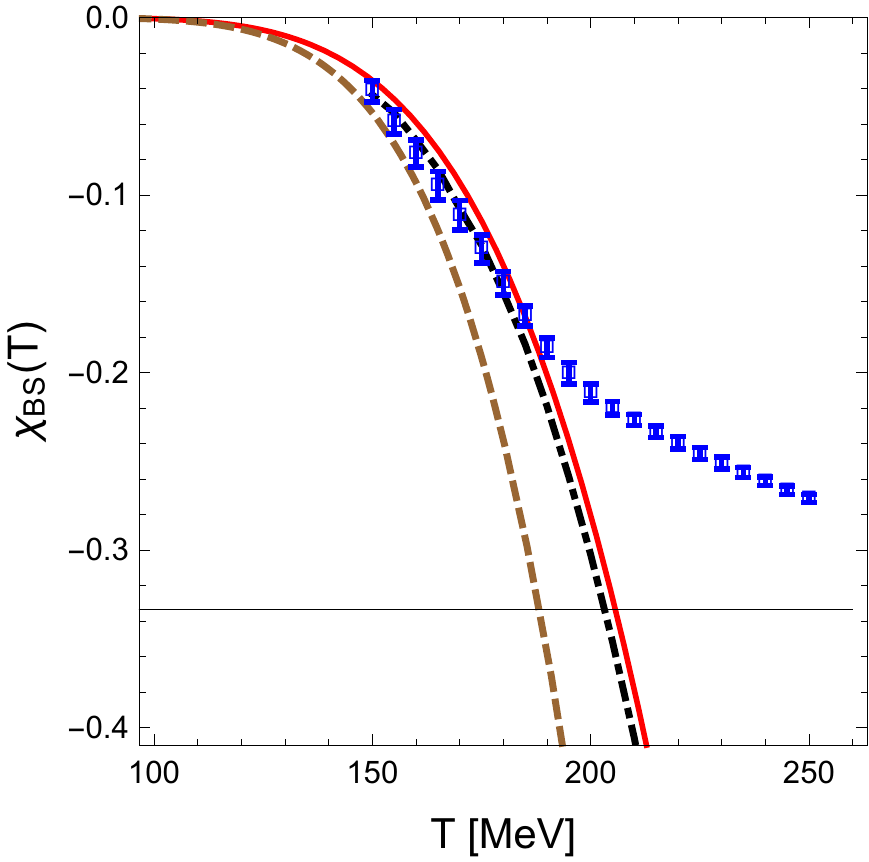,width=5.7cm,angle=0}
\epsfig{figure=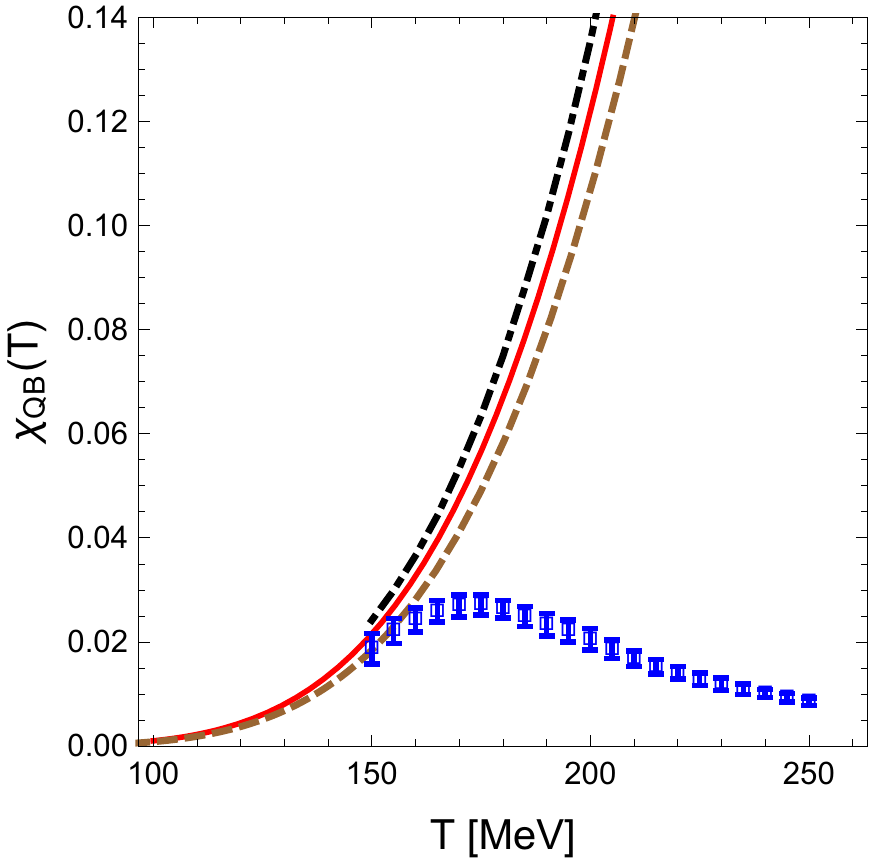,width=5.7cm,angle=0}
\epsfig{figure=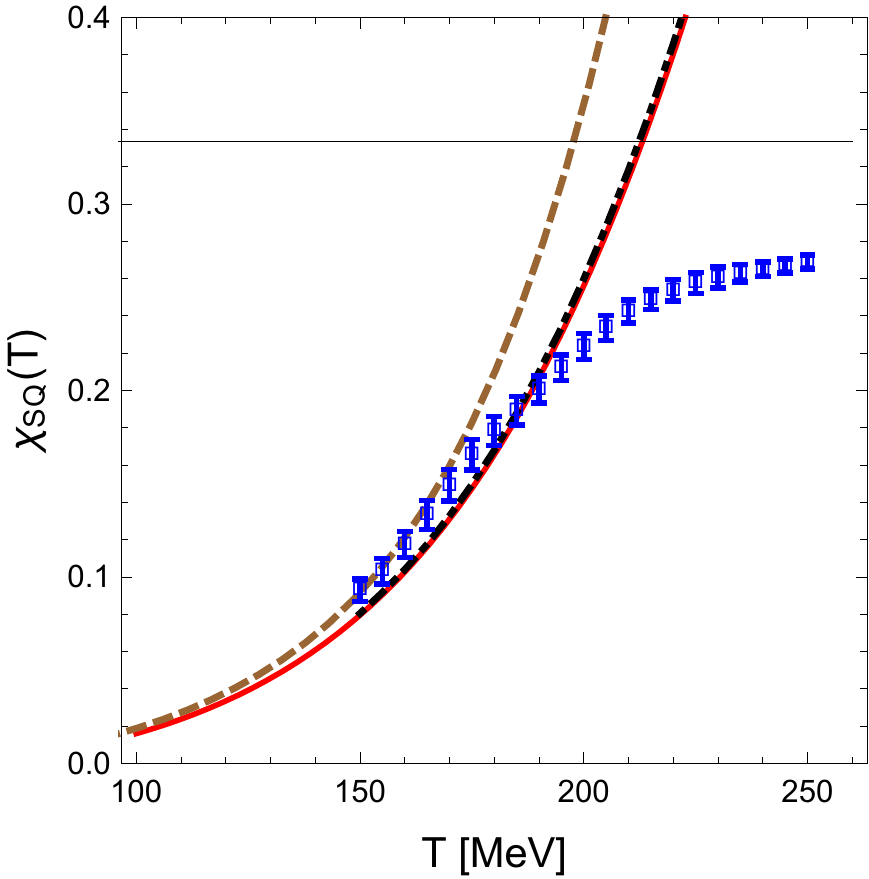,width=5.7cm,angle=0}

\caption{Baryon, charge and strangeness susceptibilities from HRG with
  the PDG, PDG($\Gamma$) and RQM spectra, compared to the lattice
  HotQCD~\cite{Bazavov:2012jq} data. WB data~\cite{Borsanyi:2011sw}
are compatible with them so they are not plotted.}
\label{fig:HRG-PDG}
\end{figure*}

Our results for the susceptibilities are depicted in
Fig.~\ref{fig:HRG-PDG} where we show the HotQCD lattice
data~\cite{Bazavov:2012jq} (the earlier WB data~\cite{Borsanyi:2011sw}
are compatible with them so they are not included in the figure to
avoid cluttering.). We compare with the standard HRG model, denoted as
PDG, the HRG including a Gaussian width profile, which we denote as
PDG ($\Gamma$), and the RQM. 

Our scheme here is to include {\it all} states from PDG, which as
mentioned are mapped into the standard quark model classification of
mesons as $\bar q q$ and baryons as $qqq$ as the only hadronic
states. This choice of states provides a visible effect in the $SB$
correlator bringing it closer to the lattice data as compared
to~\cite{ManLo:2016pgd} where only $****$ PDG states are
considered. The inclusion of width effects is also generally quite
sizeable and cannot be ignored, as it has routinely been done in many
HRG comparisons in the past (see however
\cite{Arriola:2012vk,Broniowski:2016hvt}). Nevertheless, there are other
ways to include the width profile which will somehow blur the
PDG($\Gamma$) result, and a more systematic study, perhaps including
also volume effects, would be most helpful.

The remarkable good agreement of the trace anomaly found between PDG
and RQM~\cite{Arriola:2014bfa} or the
PDG($\Gamma$)~\cite{Broniowski:2016hvt} compared with lattice QCD
results from WB~\cite{Borsanyi:2013bia} and
HotQCD~\cite{Bazavov:2014pvz} collaborations gets a bit spoiled in
terms of the considered fluctuations, where these spectra may feature
missing or exceeding states. For instance, a look at the $BB$
correlation in Fig.~\ref{fig:HRG-PDG} suggests that the RQM has too
many baryonic states but not too many charged states. Therefore, the
thermodynamic equivalence will depend on the quantum numbers,
enhancing the relevance of a fluctuation analysis, as done here, in
the discussion of quark-hadron duality.

\section{Conclusions}

In the present contribution we have revised the thermodynamical
equivalence between the PDG, RQM, and lattice QCD for temperatures
below the hadron-gas---quark-gluon-plasma crossover for the case of an
entropy shift due to a heavy quark and fluctuations via Baryon, Charge
and Strangeness susceptibilities as diagnostic tools for missing
states.

The analysis of the entropy shift due to a heavy quark suggests that
there are conventional (high mass) missing states in single charm, or
bottom hadrons ($Q \bar q$ and $Q qq$) and it looks likely that a
large number of hybrids ($Q \bar q g$) is also missing.

In the pure light $uds$ sector, our perception on the missing states may
change when finite width effects are placed into the calculation.
This effectively corresponds to redistribute the mass spectrum
weighted with an asymmetric Boltzmann factor. From that point of view
the missing states effect could also be regarded as a missing mass
effect. At this level the highest temperature of agreement for the
trace anomaly seems to be $T \lesssim 150{\rm MeV}$ between either the
HRG based on PDG, PDG ($\Gamma$) or RQM spectra and current QCD finite
temperature calculations. However, the separate analysis in terms of
$B,Q,S$ fluctuations reveals a less obvious pattern regarding the
verification of quark-hadron duality. While the HRG has arbitrated the
lattice QCD discrepancies for the trace anomaly in the past, in the
case of fluctuations we are now confronted with the opposite
situation. Lattice data agree but are not universally reproduced by
any of the three HRG realizations considered here. This may offer a
unique opportunity to refine these models including other effects and
which deserves further studies.

\begin{acknowledgments}

We thank Pok Man Lo and Michal Marczenko for useful communications.
This work is supported by Spanish Ministerio de Econom\'{\i}a y
Competitividad and European FEDER funds under contracts
FIS2014-59386-P and FPA2015-64041-C2-1-P, Junta de Andaluc\'{\i}a
grant FQM-225, and Spanish Consolider Ingenio 2010 Programme CPAN
(CSD2007-00042). W.B. is supported by the Polish National Science
Center grant 2015/19/B/ST2/00937.  The research of E.M. is supported
by the European Union under a Marie Curie Intra-European fellowship
(FP7-PEOPLE-2013-IEF) with project number PIEF-GA-2013-623006, and by
the Universidad del Pa\'{\i}s Vasco UPV/EHU, Bilbao, Spain, as a
Visiting Professor.

\end{acknowledgments}

%\bibliography{refs}

\begin{thebibliography}{10}

\bibitem{Hagedorn:1965st}
R. Hagedorn,
\newblock Nuovo Cim. Suppl. 3 (1965) 147.

\bibitem{Broniowski:2000bj}
W. Broniowski and W. Florkowski,
\newblock Phys. Lett. B490 (2000) 223, hep-ph/0004104.

\bibitem{Broniowski:2004yh}
W. Broniowski, W. Florkowski and L.{\relax Ya}. Glozman,
\newblock Phys. Rev. D70 (2004) 117503, hep-ph/0407290.

\bibitem{Olive:2016xmw}
Particle Data Group, C. Patrignani et~al.,
\newblock Chin. Phys. C40 (2016) 100001.

\bibitem{Edwards:2012fx}
Hadron Spectrum, R.G. Edwards et~al.,
\newblock Phys. Rev. D87 (2013) 054506, 1212.5236.

\bibitem{Godfrey:1985xj}
S. Godfrey and N. Isgur,
\newblock Phys. Rev. D32 (1985) 189.

\bibitem{Capstick:1986bm}
S. Capstick and N. Isgur,
\newblock Phys. Rev. D34 (1986) 2809,
\newblock [AIP Conf. Proc.132,267(1985)].

\bibitem{Broniowski:2000hd}
W. Broniowski,
\newblock {Few-quark problems. Proceedings, Mini-Workshop, Bled, Slovenia, July
  8-15, 2000}, p.~3, 2000, hep-ph/0008112.

\bibitem{Arriola:2014bfa}
E. Ruiz~Arriola, L.L. Salcedo and E. Megias,
\newblock Acta Phys. Polon. B45 (2014) 2407, 1410.3869.

\bibitem{Briceno:2015rlt}
R.A. Brice\~no et~al.,
\newblock Chin. Phys. C40 (2016) 042001, 1511.06779.

\bibitem{Majumder:2010ik}
A. Majumder and B. Muller,
\newblock Phys. Rev. Lett. 105 (2010) 252002, 1008.1747.

\bibitem{Broniowski:2016hvt}
W. Broniowski,
\newblock {Mini-Workshop Bled 2016: Quarks, Hadrons, Matter Bled, Slovenia,
  July 3-10, 2016}, 2016, 1610.09676.

\bibitem{brush1967foundations}
S.G. Brush,
\newblock Archive for History of Exact Sciences 4 (1967) 145.

\bibitem{bernoulli1738hydrodynamica}
D. Bernoulli,
\newblock Hydrodynamica sive de viribus et motibus fluidorum commentarii
  (Johann Reinhold Dulsecker, 1738).

\bibitem{Nieves:2001wt}
J. Nieves and E. Ruiz~Arriola,
\newblock Phys. Rev. D64 (2001) 116008, hep-ph/0104307.

\bibitem{GarciaRecio:2002td}
C. Garcia-Recio et~al.,
\newblock Phys. Rev. D67 (2003) 076009, hep-ph/0210311.

\bibitem{Ciulli:1975sm}
S. Ciulli, C. Pomponiu and I. Sabba~Stefanescu,
\newblock Phys. Repts.  (1975),
\newblock [Acta Phys. Austriaca Suppl.14,469(1975)].

\bibitem{Caprini:2016uxy}
I. Caprini et~al.,
\newblock Phys. Rev. D93 (2016) 076004, 1602.02062.

\bibitem{Arriola:2011en}
E. Ruiz~Arriola and W. Broniowski,
\newblock {Proceedings, Mini-Workshop on Understanding hadronic spectra: Bled,
  Slovenia, July 3-10, 2011}, pp. 7--17, 2011, 1110.2863.

\bibitem{Masjuan:2012gc}
P. Masjuan, E. Ruiz~Arriola and W. Broniowski,
\newblock Phys. Rev. D85 (2012) 094006, 1203.4782.

\bibitem{Witten:1979kh}
E. Witten,
\newblock Nucl. Phys. B160 (1979) 57.

\bibitem{Bohm:2004zi}
A.R. Bohm and Y. Sato,
\newblock Phys. Rev. D71 (2005) 085018.

\bibitem{Berggren:1968zz}
T. Berggren,
\newblock Nucl. Phys. A109 (1968) 265.

\bibitem{Masjuan:2012sk}
P. Masjuan, E. Ruiz~Arriola and W. Broniowski,
\newblock Phys. Rev. D87 (2013) 014005, 1210.0760.

\bibitem{Briceno:2014tqa}
R.A. Brice\~no, Z. Davoudi and T.C. Luu,
\newblock J. Phys. G42 (2015) 023101, 1406.5673.

\bibitem{Dashen:1969ep}
R. Dashen, S.K. Ma and H.J. Bernstein,
\newblock Phys. Rev. 187 (1969) 345.

\bibitem{Dashen:1974jw}
R.F. Dashen and R. Rajaraman,
\newblock Phys. Rev. D10 (1974) 694.

\bibitem{Dashen:1974yy}
R.F. Dashen and R. Rajaraman,
\newblock Phys. Rev. D10 (1974) 708.

\bibitem{Venugopalan:1992hy}
R. Venugopalan and M. Prakash,
\newblock Nucl. Phys. A546 (1992) 718.

\bibitem{Broniowski:2015oha}
W. Broniowski, F. Giacosa and V. Begun,
\newblock Phys. Rev. C92 (2015) 034905, 1506.01260.

\bibitem{Arriola:2015gra}
E. Ruiz~Arriola, L.L. Salcedo and E. Megias,
\newblock Acta Phys. Polon. Supp. 8 (2015) 439, 1505.02922.

\bibitem{Arriola:2012vk}
E. Ruiz~Arriola, W. Broniowski and P. Masjuan,
\newblock (2012), 1210.7153,
\newblock [Acta Phys. Polon. Supp.6,95(2013)].

\bibitem{Megias:2004hj}
E. Megias, E. Ruiz~Arriola and L.L. Salcedo,
\newblock Phys. Rev. D74 (2006) 065005, hep-ph/0412308.

\bibitem{Megias:2006bn}
E. Megias, E. Ruiz~Arriola and L.L. Salcedo,
\newblock Phys. Rev. D74 (2006) 114014, hep-ph/0607338.

\bibitem{Fukushima:2003fw}
K. Fukushima,
\newblock Phys. Lett. B591 (2004) 277, hep-ph/0310121.

\bibitem{Ratti:2005jh}
C. Ratti, M.A. Thaler and W. Weise,
\newblock Phys. Rev. D73 (2006) 014019, hep-ph/0506234.

\bibitem{RuizArriola:2012wd}
E. Ruiz~Arriola, E. Megias and L.L. Salcedo,
\newblock AIP Conf. Proc. 1520 (2013) 185, 1207.4875.

\bibitem{Megias:2013xaa}
E. Megias, E. Ruiz~Arriola and L.L. Salcedo,
\newblock Phys. Rev. D89 (2014) 076006, 1311.2814.

\bibitem{Borsanyi:2013bia}
S. Borsanyi et~al.,
\newblock Phys. Lett. B730 (2014) 99, 1309.5258.

\bibitem{Bazavov:2014pvz}
HotQCD, A. Bazavov et~al.,
\newblock Phys. Rev. D90 (2014) 094503, 1407.6387.

\bibitem{Megias:2012kb}
E. Megias, E. Ruiz~Arriola and L.L. Salcedo,
\newblock Phys. Rev. Lett. 109 (2012) 151601, 1204.2424.

\bibitem{Megias:2012hk}
E. Megias, E. Ruiz~Arriola and L.L. Salcedo,
\newblock (2012), 1207.7287,
\newblock [Nucl. Phys. Proc. Suppl.234,313(2013)].

\bibitem{Megias:2016bhk}
E. Megias, E. Ruiz~Arriola and L.L. Salcedo,
\newblock Acta Phys. Polon. Supp. 9 (2016) 401, 1605.04453.

\bibitem{RuizArriola:2016ift}
E. Ruiz~Arriola, L.L. Salcedo and E. Megias,
\newblock {12th Conference on Quark Confinement and the Hadron Spectrum
  (Confinement XII) Thessaloniki, Greece, August 28-September 2, 2016}, 2016,
  1611.03255.

\bibitem{Megias:2016onb}
E. Megias, E. Ruiz~Arriola and L.L. Salcedo,
\newblock Phys. Rev. D94 (2016) 096010, 1603.04642.

\bibitem{Bazavov:2016uvm}
A. Bazavov et~al.,
\newblock Phys. Rev. D93 (2016) 114502, 1603.06637.

\bibitem{Jeon:1999gr}
S. Jeon and V. Koch,
\newblock Phys. Rev. Lett. 83 (1999) 5435, nucl-th/9906074.

\bibitem{Jeon:2000wg}
S. Jeon and V. Koch,
\newblock Phys. Rev. Lett. 85 (2000) 2076, hep-ph/0003168.

\bibitem{Koch:2008ia}
V. Koch,
\newblock {Hadronic Fluctuations and Correlations},
\newblock Chapter of the book Relativistic Heavy Ion Physics, R. Stock (Ed.),
  Springer, Heidelberg, 2010, p. 626-652.

\bibitem{Asakawa:2015ybt}
M. Asakawa and M. Kitazawa,
\newblock Prog. Part. Nucl. Phys. 90 (2016) 299, 1512.05038.

\bibitem{Braun-Munzinger:2015hba}
P. Braun-Munzinger et~al.,
\newblock Phys. Rept. 621 (2016) 76, 1510.00442.

\bibitem{Bazavov:2014xya}
A. Bazavov et~al.,
\newblock Phys. Rev. Lett. 113 (2014) 072001, 1404.6511.

\bibitem{ManLo:2016pgd}
P. Man~Lo et~al.,
\newblock Eur. Phys. J. A52 (2016) 235.

\bibitem{Borsanyi:2011sw}
S. Borsanyi et~al.,
\newblock JHEP 01 (2012) 138, 1112.4416.

\bibitem{Bazavov:2012jq}
HotQCD, A. Bazavov et~al.,
\newblock Phys. Rev. D86 (2012) 034509, 1203.0784.

\bibitem{Ejiri:2005wq}
S. Ejiri, F. Karsch and K. Redlich,
\newblock Phys. Lett. B633 (2006) 275, hep-ph/0509051.

\bibitem{Bellwied:2015lba}
R. Bellwied et~al.,
\newblock Phys. Rev. D92 (2015) 114505, 1507.04627.

\end{thebibliography}
%\bibliographystyle{h-elsevier}

\end{document}